# APCs and citation impact of Gold OA articles authored by Ukrainian scholars before and during Russia's full-scale war against Ukraine (2020-2023)

**Corresponding author:**
Myroslava Hladchenko
hladchenkom@gmail.com

## Abstract
This study first examines how APC expenditures, authorship patterns, and publishing venues of Ukrainian scholars changed between the pre-war (2020–2021) and wartime (2022–2023) periods. Second, it explores the extent to which APC levels are associated with the field-normalised citation impact (FNCI) of Gold Open Access articles authored by Ukrainian scholars. Statistical analysis revealed a small but significant correlation between APC amounts and citation impact, though the effect size was minimal, suggesting higher APCs did not substantially boost citations. APC waivers offered by major publishers such as Springer and Elsevier since 2022 resulted in only a slightly increased number of articles authored solely by Ukrainian scholars. Despite these waivers, MDPI and Aluna maintained the largest shares. Between 2020 and 2023, the number of articles authored solely by Ukrainian scholars in foreign journals fell by 25.7%, and total APC spending declined by 24.6%, from €1.24 million to €0.93 million. Medicine accounted for the largest share of both articles and APC expenditure, with the majority published in Aluna journals.

## Introduction
Open Access (OA) publishing was introduced to address issues of accessibility, affordability, and equity in scholarly communication. While OA publishing, especially models based on article processing charges (APCs), has removed barriers to accessing research, it has simultaneously created new obstacles for publishing (Ellers et al., 2017). This has, arguably, worsened the situation (Anderson, 2023; Scott, 2018). The challenge of covering APCs is not confined to developing countries; institutions in developed countries such as the UK also face financial constraints (Barr, 2025). The financial burden of APCs on institutions and scholars has far-reaching implications, negatively affecting both the scientific community and individual researchers (Asai 2024; Halevi and Walsh 2021; Haustein et al. 2024; Alonso-Álvarez et al. 2024; Nejadghanbar & Hu, 2022).

Moreover, although the APC model aims to make research freely available to the public, it can be exploited by profit-driven publishers that fail to uphold rigorous peer review standards (Bohannon, 2013). Striving for promotion and aiming for better research evaluation results, scholars can prioritise quick publishing over quality of research (Csomós and Farkas 2023). The relationship between APCs, citation impact, and publication quality remains contested. While Björk and Solomon (2015) and Maddi and Sapinho (2022) found a moderate correlation between APCs and citation impact, the overall evidence remains inconclusive (Pollock and Michael 2019).

This study addresses challenges of APC-based Gold OA publishing in Ukraine, a country experiencing Russia's full-scale invasion since 2022. The paradox of war is that, while life-threatening conditions, shelling, blackouts and economic recession are a part of everyday reality (OECD, 2022), work obligations and responsibilities remain the same as they were in pre-war life. This affects scholars as well (UNESCO, 2024). In 2022-2024, 1,518 Ukrainian scholars have joined the army (Cabinet of Ministers of Ukraine 2024). A total of 146 Ukrainian scholars were killed, comprising individuals who served in the army as well as those killed in Russian missile attacks (Novynarnia, 2024). Those not serving in the army, like other Ukrainians, face regular shelling and power blackouts. Not to mention the economic crisis caused by Ukraine allocating all resources to the defence sector (BBC, 2024). Despite Russia's full-scale invasion, Ukrainian scholars continue to publish, which deserves respect.

In response to Russia's full-scale war against Ukraine, several international publishers—such as Wiley, IOP Publishing, PLOS, SAGE, Springer, Taylor & Francis, and Bentham Science

Publishers—have introduced full APC waivers for corresponding authors affiliated with Ukrainian institutions. Polish ALUNA Publishing House have provided a partial discount by reducing the APC from €600 to €350.

This study addresses two research questions:

1.How did APC expenditures, authorship patterns, and publishing venues of Ukrainian scholars change between the pre-war (2020–2021) and wartime (2022–2023) periods?

2. To what extent are APC levels associated with the field-normalised citation impact (FNCI) of Gold Open Access articles authored by Ukrainian scholars?

Data were taken from the CWTS in-house Scopus database and consist of articles published by Ukrainian scholars before (2020-2021) and after the start (2022-2023) of Russia's full-scale war on Ukraine.

**APCs and citation impact**

Open access was intended to solve the problems of accessibility, affordability, and equity (Anderson, 2023). Nation-states and research institutions viewed OA publishing as an alternative to the subscription model, aiming not only to make research outputs publicly accessible but also to reduce subscription costs. However, as open access was implemented mainly through the APC model in Gold OA and Hybrid journals, it created inequalities and unaffordability in publishing for scholars (Brainard 2024; Demeter and Istratii 2020; Olejniczak and Wilson 2020; Siler et al. 2018). APCs are a large burden for the national science systems, as well as for institutions (Zhang et al. 2022; Pinfield et al. 2016; Johnson et al. 2016). Covering APCs is challenging even for developed countries. In 2025, three UK universities ended their Read-and-Publish Deals with Elsevier due to high costs (Barr 2025). Alonso-Álvarez et al. (2024) revealed that APC spending from Spanish publicly funded projects between 2013 and 2019 equaled the entire budget allocated to the field of psychology during the same period. They argue that these findings highlight the necessity to reflect on alternative uses of this money, both related to pure scientific, such as personnel hiring or equipment, or editorial purposes, such as the development of alternative publishing models.

For publishers, OA publishing based on APCs turned out to be an even more profitable business model than subscription, which in turn, increased the costs incurred by institutions (Maddi 2020; Maddi and Sapinho 2021). Between 2016 and 2020, traditional publishers—led by Springer Nature, followed by Elsevier and Wiley—doubled the number of articles in Gold OA journals, while APC revenues tripled. Haustein et al. (2024) estimated that a total of $8.349 billion was spent on APCs to Elsevier, Frontiers, MDPI, PLOS, Springer Nature, and Wiley between 2019 and 2022. In 2023, MDPI ($681.6 million), Elsevier ($582.8 million) and Springer Nature ($546.6 million) generated the most revenue with APCs.

Schönfelder's (2020) analysis of UK research output found that most APC-funded articles were published by Elsevier, Springer Nature, and Wiley-Blackwell, with 4% of all articles appearing in the pure open-access megajournal PLOS ONE. Shu and Larivière (2024) found similar global trends for 2008–2020, noting a decline in PLOS's share and a rise in MDPI's. By 2020, MDPI had become the world's largest OA publisher. Compared to the Big Five publishers (Springer Nature, Wiley, Elsevier, Taylor & Francis, Sage), MDPI journals offer faster turnaround times and higher acceptance rates (Hanson et al. 2024; Csomós and Farkas, 2023). However, MDPI journals exhibit higher rates of self-citation and cross-citation within their own portfolio (Seeber et al. 2024; Hanson et al. 2024). Reflecting these concerns, Finland's Publication Forum downgraded 193 MDPI journals to its lowest level 0 rating in 2024, while only 16 retained level 1 (Publication Forum, 2024). Western European countries predominantly publish with the Big Five, whereas Central and Eastern European countries rely heavily on MDPI journals (Sasvári and Urbanovics 2023; Cernat 2024; Csomós and Farkas 2023).

Another negative consequence of the APC-based open access publishing model is the decline in publication quality, as some high-quality journals may become more lenient and accept lower-quality articles than they would under subscription models (Van Vlokhoven 2019). Additionally, the APC model—while intended to make research publicly accessible—can be exploited by publishers prioritizing profit over rigorous peer review (Bohannon 2013). This practice risks undermining the integrity and reliability of scientific knowledge (Seeber et al. 2024).

The APC-based Gold OA publishing model is more established in STEM than social sciences and humanities (Schönfelder 2020). Zhang et al. (2020) attribute this to social sciences and humanities having less funding than health and STEM sciences. Diamond OA predominates in the social sciences and humanities, which have a higher number of Diamond OA journals than other disciplines (Bosman et al. 2021; Taubert et al. 2024; van Bellen and Céspedes 2024).

The findings on the relationship between citation impact, APC and quality of publications are rather controversial (Björk and Solomon 2015; Pollock and Michael 2019). APCs paid to open-access journals were found to be correlated with higher journal citation impact (Romeu et al. 2014; Schönfelder 2020; Asai 2019). However, Björk and Solomon (2015) and Maddi and Sapinho (2022) revealed a moderate correlation between APCs and citation impact. Paying high APCs does not necessarily increase the impact of publications because publishers with the highest APCs are not necessarily the best in terms of impact.

## Open Science in Ukraine

After the fall of the Soviet Union in 1991, Ukraine maintained a division between primarily teaching-oriented higher education institutions and research institutes under the National Academy of Sciences of Ukraine (NASU). The Academy comprises research institutes organised into fourteen disciplinary sections, with physical and technical sciences dominating in terms of researchers and institutes compared to biological sciences. In addition to NASU, sectoral academies were established after 1991, including medical, educational, agricultural, and law academies

Higher education institutions are also authorised to conduct fundamental and applied research. In 2014, the maximum teaching hours for academics were reduced from 900 to 600 per year (Parliament of Ukraine, 2014) to allow more time for research. After 1991, publication requirements for doctoral degrees and promotion to professor included articles published only in Ukrainian journals. Many Ukrainian journals adopted open access model charging APCs, which scholars typically covered themselves as institutions did not allocate funding for APCs.

In 2013, the Ministry of Education introduced a requirement for doctoral candidates to publish in Ukrainian journals indexed by Scopus and Web of Science (WoS) (Ministry of Education and Science, Youth and Sport of Ukraine 2012). Initially, the focus was on Ukrainian Scopus-indexed journals. In subsequent years, publications in both Ukrainian and international Scopus- and WoS-indexed journals became part of the licensing criteria for study programmes (Cabinet of Ministers of Ukraine, 2015), requirements for academic titles of associate professor and professor (Ministry of Education and Science of Ukraine, 2016), and criteria for the research assessment of higher education institutions (Cabinet of Ministers of Ukraine 2017). Since Ukrainian scholars often paid for publishing in Ukrainian journals, most assumed that international Scopus- and WoS-indexed journals also charge APCs.

Previous studies have shown that the introduction of publication requirements emphasising Scopus-indexed journals led to an increase in the number of such articles published by Ukrainian scholars (Hladchenko 2022). However, quantity was prioritised over quality, and Ukrainian academics also published extensively in local Scopus-indexed journals (Nazarovets 2020; 2022). In response to Russia's full-scale war against Ukraine, international publishers introduced waivers for Gold OA journals for corresponding authors affiliated with Ukrainian institutions. These include Wiley, IOP Publishing, PLOS, Sage, Springer, Taylor & Francis, and Bentham Science Publishers. Polish journals such as *Archives of Materials Science and Engineering* and *Journal of Achievements in Materials and Manufacturing Engineering* also provided waivers for Ukrainian scholars. Other publishers offered discounts—for example, the Polish ALUNA Publishing House, which normally charges €600 for foreign authors, has reduced fees to €350 for Ukrainian authors since 2022. De Gruyter and Engineered Science Publisher offer a 50% discount for Ukraine as a lower middle-income country. MDPI has not announced waivers for Ukrainian scholars; however, it provides reviewer vouchers that partially or fully cover APCs, and guest editors of special issues publish free of charge.

## Materials and methods

### *Materials*

First, articles published between 2020 and 2023 with at least one author (co-)affiliated with a Ukrainian institution were extracted from the CWTS in-house database. Based on Ukrainian affiliations, the dataset includes 61,106 articles, categorised as follows: 17,554 authored by scholars affiliated with the National Academy of Sciences of Ukraine (NASU), 45,496 by those affiliated with universities, and 4,099 by scholars from other institutions.

Scopus classifies open access (OA) into four types: Gold OA, Hybrid Gold OA, Green OA, and Bronze OA. However, Scopus does not distinguish whether Gold OA journals charge article processing charges (APCs). Gold OA journals that do not charge APCs are referred to as Diamond OA. To differentiate between Gold OA (charging APCs) and Diamond OA (no APC), data from the Directory of Open Access Journals (DOAJ) were used. For journals not indexed in DOAJ, access models were examined manually on the journals' websites. Journals that publish open access and either explicitly state they do not charge APCs or have no APC information were classified as Diamond OA. Journals operating under the Subscribe to Open (S2O) model, which provide open access without charging APCs, were also classified as Diamond OA (e.g., *Journal of Fractal Geometry*).

Using the Directory of Open Access Journals (DOAJ) list and manual examination of journal websites resulted in the recategorisation of some articles initially classified by Scopus as Bronze OA, Green OA, Hybrid Gold OA, or closed access. For example, a review of articles classified as Hybrid Gold OA revealed that some were actually published in Gold OA or Diamond OA journals. Specifically, *Slavica Slovaca*, *European Studies: The Review of European Law, Economics and Politics*, *Journal of Ethnology and Folkloristics*, and *ESAIM – Control, Optimisation and Calculus of Variations* were recategorised as Diamond OA. Meanwhile, *Acta Archaeologica Lodziensia*, *International Journal of Information Systems in the Service Sector* (since January 2023), *Review of Economics and Finance*, *Statistics, Optimization and Information Computing*, *Czasopismo Geograficzne* (since 2022), *Contemporary Mathematics* (Singapore), and *International Journal of Statistics in Medical Research* were recategorised as Gold OA.

Manual examination revealed several discrepancies in Scopus data: some journals do not report their Gold OA status to Scopus (e.g., *International Journal of Entrepreneurship*, *International Journal of Innovation, Creativity and Change*, *International Journal of Learning, Teaching and Education Research*); some journals transitioned to Gold OA between 2021 and 2023 but this change is not reflected in Scopus metadata, except for Springer and Elsevier journals (e.g., *Colloids and Interface Science Communications*); and some articles linked as open access in Scopus lack formal OA status, such as those in *News of the National Academy of Sciences of the Republic of Kazakhstan* and *Series of Geology and Technical Sciences*.

A dataset of 26,669 Gold OA articles authored by Ukrainian scholars was compiled, including 13,774 articles published in foreign journals and 12,895 in Ukrainian journals. To identify Ukrainian journals, the list compiled by Hladchenko and Moed (2022) was used, supplemented with data from the DOAJ and manual verification. Journals were categorised as Ukrainian or foreign based on their country of publication. Journals published by Springer but managed by Ukrainian editorial boards or institutions were also considered Ukrainian for this study. For example, the *Journal of Mathematical Sciences* includes Series A and B, with Series B comprising English-language translations from 12 Russian- and Ukrainian-language source journals. The present analysis focuses on articles published in foreign journals.

The DOAJ list was used to extract data on APCs. Missing APC data were gathered through manual examination of journal websites. APC amounts were converted into euros to ensure comparability. Full waivers announced by publishers publicly for Ukrainian corresponding authors after 2022, were incorporated in a separate analysis comparing APCs with and without waivers, while acknowledging that the actual uptake of waivers cannot be verified at the article level.

***Methods***

Using authors' country affiliation data, the following authorship patterns were identified: (1) internationally co-authored articles led by international corresponding authors (INTER–INTER);
(2) internationally co-authored articles led by Ukrainian corresponding authors (INTER–UKR);
(3) internationally co-authored articles with corresponding authors holding both Ukrainian and foreign affiliations (INTER–UKRCOAF);
(4) articles authored solely by Ukrainian scholars (UKR–UKR);
(5) articles primarily authored by Ukrainian scholars with corresponding authors holding both Ukrainian and foreign affiliations (UKR–UKRCOAF).

This classification helps to analyse how Ukrainian scholars engaged in international collaboration and maintained leadership roles in foreign Gold OA publications, particularly during the ongoing war and in the context of APC waivers provided to scholars affiliated with Ukrainian institutions serving as corresponding authors.

Citations to articles in the dataset were extracted from CWTS Scopus in-house database released in March 2025. CWTS Scopus in house database is updated once a year. As articles in the dataset were published in 2020-2023, citations from 2025 provide the largest possible citation window from available data (Abramo et al., 2011). Disciplines differ in publication, citation, and authorship practices, as well as publications published earlier have citation advantages. To overcome these discripancies and ensure comparapitily of articles from different disciplines and published in different years, it is standard in bibliometrics to normalise citations (Bornmann & Wohlrabe, 2019). The recommendation to use normalised bibliometric indicators instead of bare citation counts is one of the ten guiding principles for research metrics listed in the Leiden manifesto (Hicks et al. 2015; Wilsdon et al. 2015).

Field-normalised citation impact (FNCI) for each article in the dataset was calculated in CWTS Scopus in-house database. Citations to each article were divided by average global number of citation to articles published in the same discipline in the same year. As such, citations to each article were normalised by discipline and year of publication. The number of citations for each article was normalised by dividing it by the global average number of citations for articles in the same discipline, publication year, based on the Scopus subject classification system. Articles classified by Scopus in more than one discipline were fractionalised among these disciplines.

To examine systematic differences in citation impact, three linear regression models were estimated. Model 1 used log-transformed normalized citation impact as the dependent variable (log(FNCI + 1)). The log transformation was applied to reduce the influence of extreme citation values and to account for the highly skewed distribution of citation counts, following established bibliometric practice (Thelwall & Wilson, 2014). The model examined the effects of publication period (2020–2021 vs. 2022–2023) and authorship type. Model 2 assessed the association between article processing charges (APCs, per €100) and field-normalized citation impact (FNCI). The model included main effects for APCs, authorship type, and publication period (2020–2021 vs. 2022–2023), as well as all two- and three-way interaction terms (APC × authorship type × period). FNCI was log-transformed (log(FNCI + 1)) to improve the normality of residuals. Model 3 examined the relationship between citation impact and publication characteristics, predicting log-transformed FNCI (log(FNCI + 1)) as a function of publication period, publisher, and authorship type, including their interaction terms. This model explained 28.2% of the variance in log(FNCI + 1) (adjusted $R^2 = 0.282$).

## Results

### *Gold OA: national perspective*

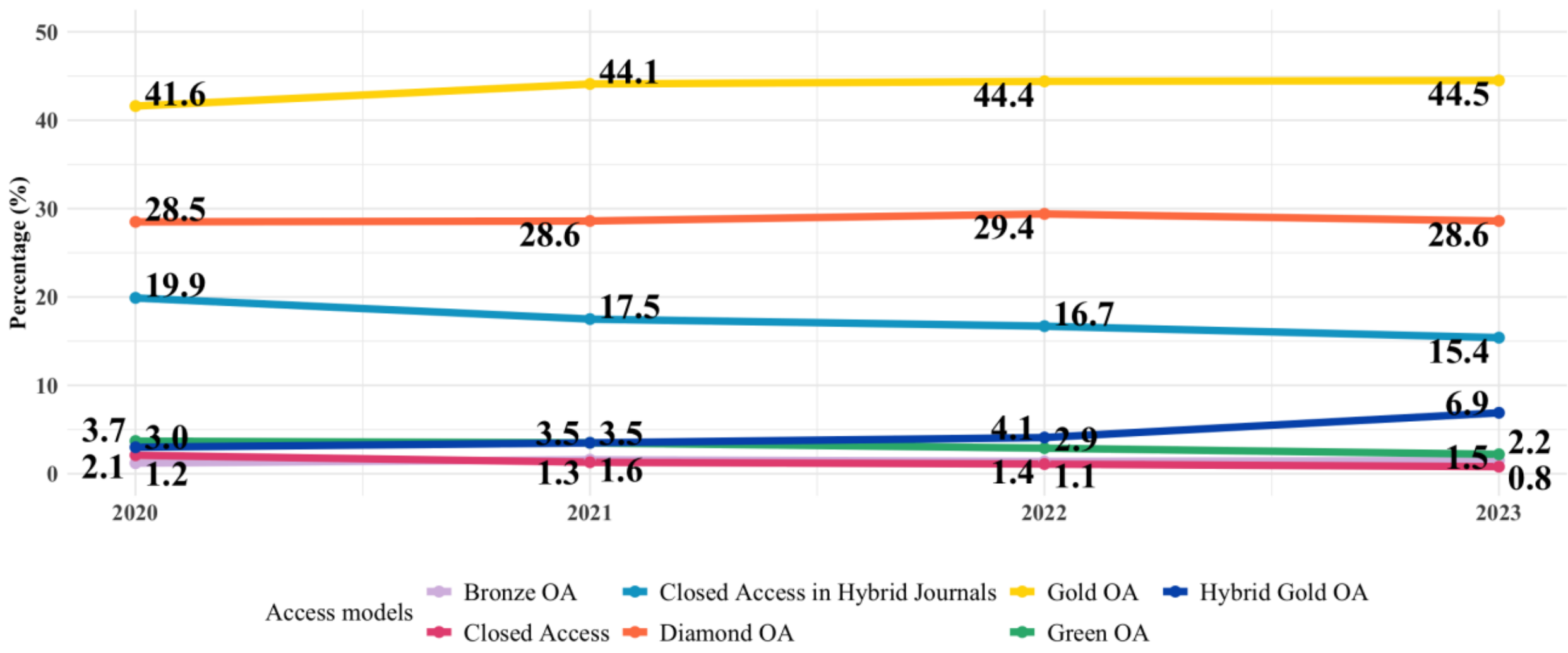

Figure 1. Distribution of articles across access models.

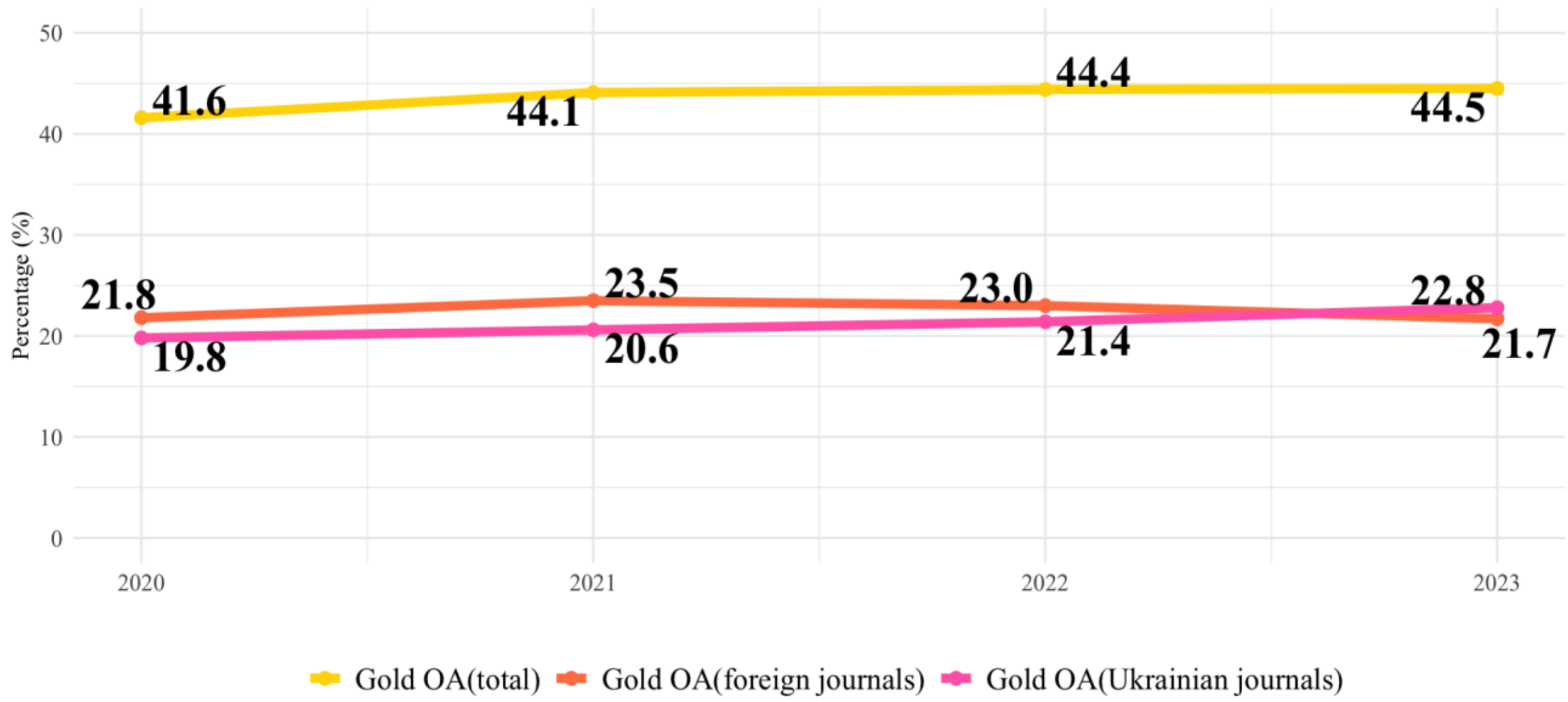

Figure 2. Distribution of Gold OA articles in foreign and Ukrainian journals.

Figure 1 shows that Gold OA articles made up the largest share of the total research output authored by Ukrainian scholars. Between 2020 and 2023, their share increased from 41.6% to 44.5%. Figure 2 shows that, between 2020 and 2023, Gold OA articles were distributed almost evenly between foreign and Ukrainian journals. During this period, the share of Gold OA in foreign journals remained virtually unchanged, slipping slightly after 2021, while the share in Ukrainian journals rose from 19.8% to 22.8%. This shift can be attributed to the financial constraints faced by Ukrainian scholars, including the high cost of APCs in many foreign journals and the economic hardships caused by the war, which made lower-cost or discounted domestic publishing options more viable.

### *Authorship types in Gold OA articles in foreign journals*

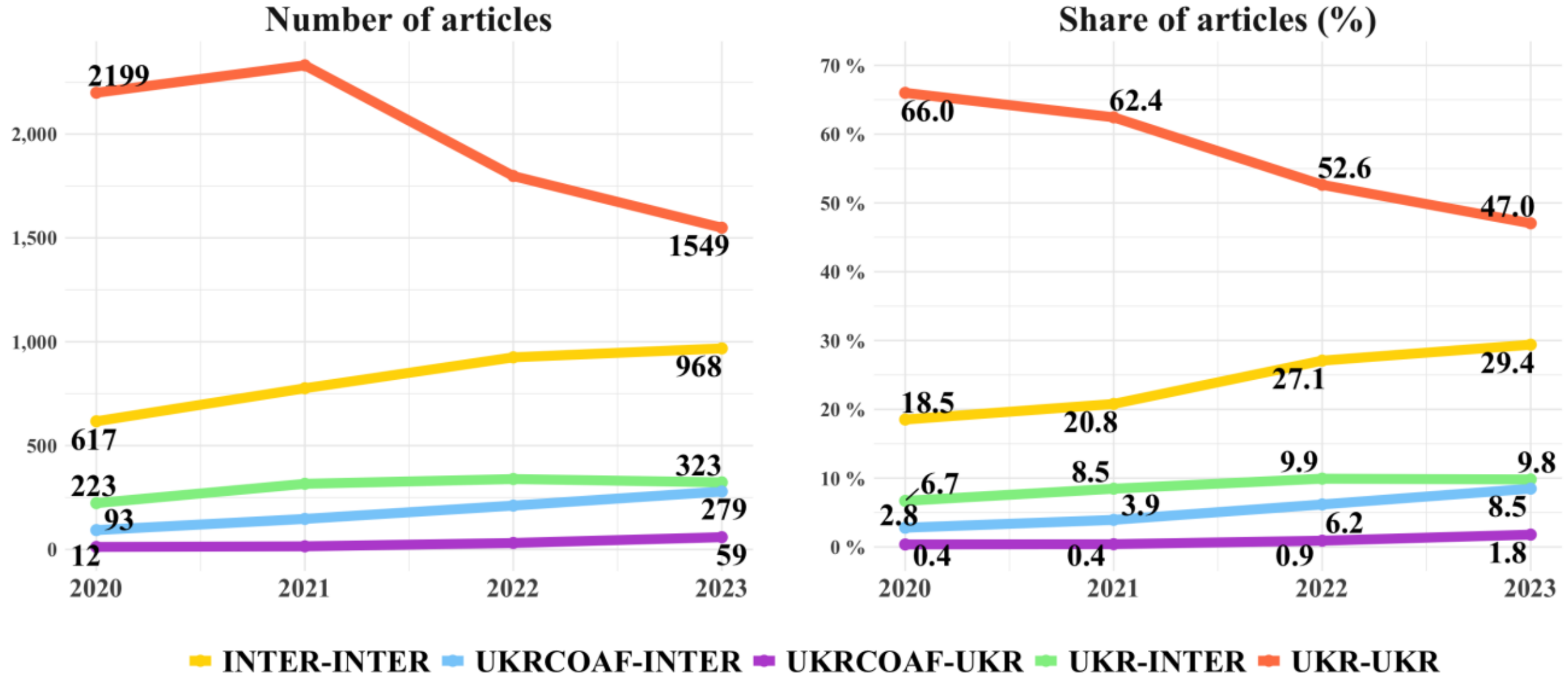

Figure 3 Distribution of Gold OA across authorship types.

Figure 3 illustrates the distribution of Gold Open Access (GOA) articles across different authorship types, defined by the country affiliation of authors and the corresponding author.
Between 2020 and 2023, purely Ukrainian publications (UKR-UKR) decline from 66.0% to 47.0% of total output. This decline is accompanied by increases in international-only articles (INTER-INTER, 18.5% → 29.4%) and in collaborative categories involving Ukrainian authors with international co-authors, including UKR-INTER (≈7–10%) and dual-affiliation corresponding authors (UKRCOAF-INTER, 2.8% → 8.5%). The pattern suggests that as local-only research decreases, Ukrainian scientists increasingly engage in international collaborations. The share of GOA articles in foreign journals with unknown corresponding author decreased from 5.6% (188) to 3.5% (115).

### *Citation impact of Gold OA in foreign journals across authorship types*

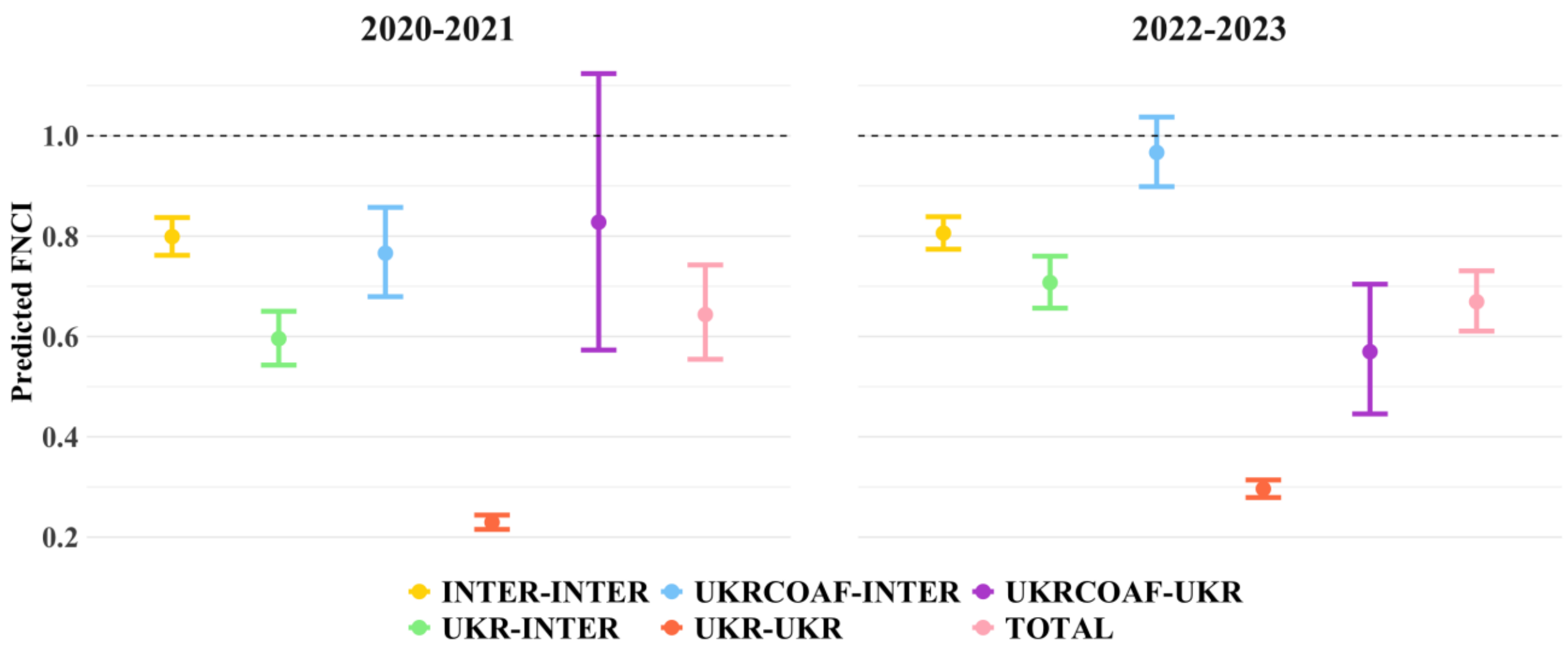

Figure 4. Predicted absolute FNCI of Gold OA in foreign journals across authorship types.

Model 1 is a linear regression model in which the log-transformed normalized citation impact (log(FNCI + 1)) was used as the dependent variable. The model was designed to examine the effects of publication period (2020–2021 vs. 2022–2023) and authorship type.

Predicted absolute FNCI values from Model 1 are presented in Figure 4, while the model summary and predicted FNCI values are reported in Tables 1 and 2 in the Appendix. In both periods, articles with only Ukrainian affiliations (UKR-UKR) show the lowest predicted impact, increasing from about 0.23 (95% CI: 0.22–0.24) in 2020–2021 to 0.30 (95% CI: 0.28–0.31) in 2022–2023. In contrast, articles with only international affiliations (INTER-INTER) consistently display the highest predicted FNCI, rising slightly from approximately 0.79 (95% CI: 0.76–0.83) to 0.84 (95% CI: 0.81–0.87).

The remaining collaboration types fall between these two groups. Articles with foreign co-affiliation for the corresponding author and international co-authors (UKRCOAF-INTER) are associated with higher predicted FNCI, 0.77 (95% CI: 0.68–0.86) in 2020–2021 and 0.97 (95% CI: 0.90–1.04) in 2022–2023, compared with mixed Ukrainian–international collaborations without corresponding author co-affiliation (UKR-INTER), 0.60 (95% CI: 0.54–0.65) in 2020–2021 and 0.71 (95% CI: 0.66–0.76) in 2022–2023. UKRCOAF-UKR article also lie within this intermediate range, although estimates are more uncertain due to fewer observations (0.83; 95% CI: 0.57–1.12 in 2020–2021 and 0.57; 95% CI: 0.45–0.70 in 2022–2023). Overall, these results reveal a consistent gradient in predicted citation impact, with the higher concentration of international affiliation for corresponding author and co-authors being associated with higher citation impact.

The overall predicted FNCI for all articles (TOTAL) increased modestly from 0.64 (95% CI: 0.55–0.74) in 2020–2021 to 0.67 (95% CI: 0.61–0.73) in 2022–2023.

### *APCs (without and with waivers) of Gold OA across authorship types*

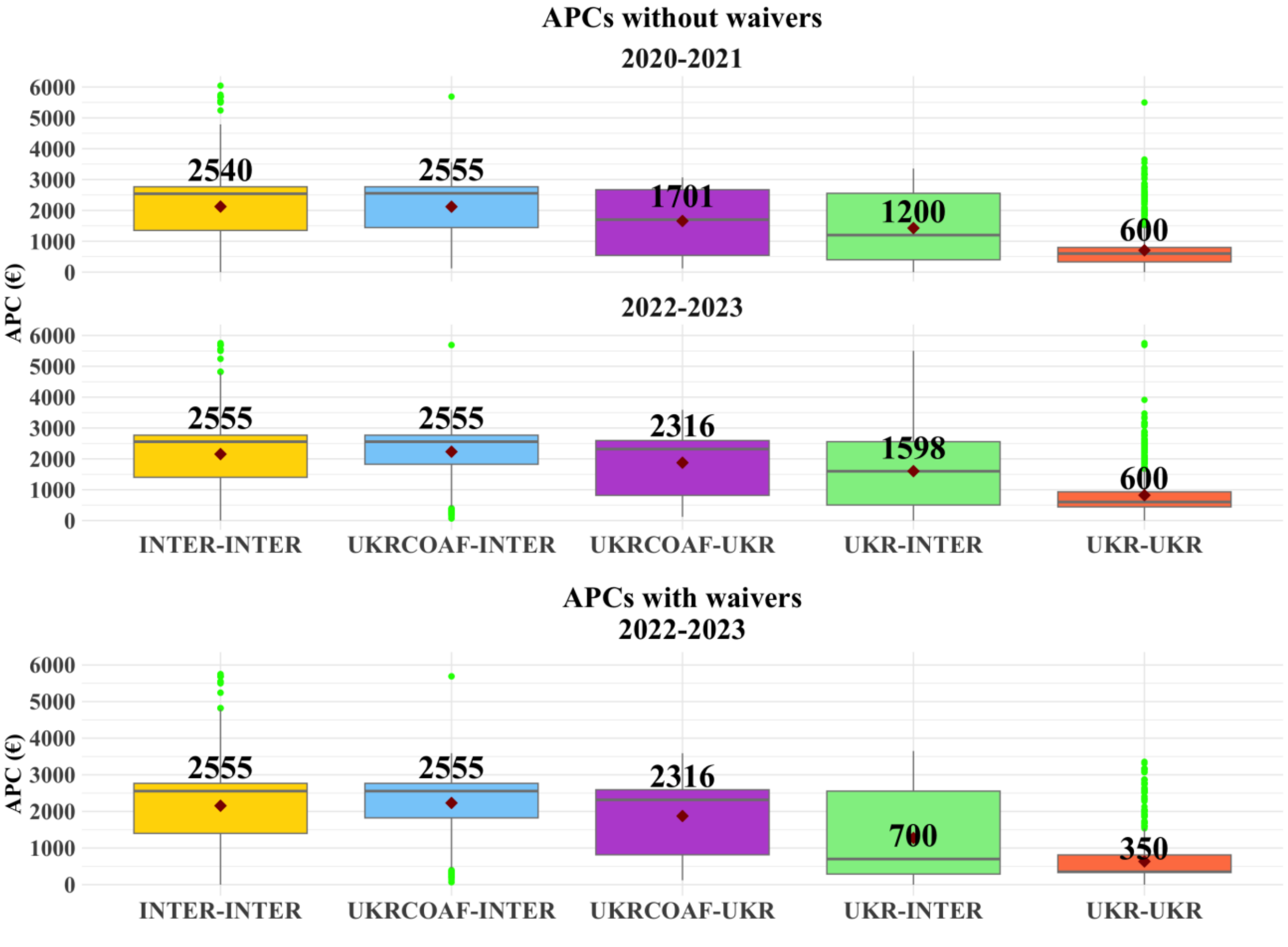

Figure 5. Median APCs of Gold OA across authorship types without and with waivers.

Figure 5 compares APCs (without and with waivers) of Gold OA across authorship types. It shows that in both periods, median APCs for internationally co-authored articles led by international scholars (INTER-INTER) (€2,540 in 2020–2021 and €2,555 in 2022–2023) and for articles led by Ukrainian scholars with foreign co-affiliations (UKRCOAF-INTER and UKRCOAF-UKR) were at a similar level, approximately €2,300–2,500. These were substantially higher than APCs for internationally co-authored articles led by Ukrainian scholars (UKR-INTER, ~€1,500 in 2020–2021). Articles authored solely by Ukrainian scholars (UKR-UKR) had the lowest median APCs in both periods (€700 in 2020–2021 and €350 in 2022–2023).

In 2022–2023, due to publisher waivers, the nominal median APC for UKR-INTER articles rose from €1,200 to €1,598, but the effective (factual) APC decreased to €700. This reduction was driven by waivers provided by Elsevier, Springer, and Wiley. Similarly, the factual APC for UKR-UKR articles decreased from €600 to €350, reflecting waivers offered by Aluna Publishing House.

Table 1 Baseline FNCI and estimated increase per €100 APC by authorship pattern and period

| **Period** | **Authorship** | **Baseline FNCI** | **ΔFNCI per €100 APC** | **CI lower** | **CI upper** |
|---|---|---|---|---|---|
| 2020-2021 | INTER-INTER | 0,350 | 0,018 | -0,088 | 0,125 |
| | UKRCOAF-INTER | 0,261 | 0,024 | -0,238 | 0,285 |
| | UKRCOAF-UKR | 0,499 | 0,019 | -0,725 | 0,756 |
| | UKR-INTER | 0,355 | 0,017 | -0,110 | 0,145 |
| | UKR-UKR | 0,228 | 0,004 | -0,033 | 0,041 |
| 2022-2023 | INTER-INTER | 0,377 | 0,017 | -0,080 | 0,115 |
| | UKRCOAF-INTER | 0,754 | 0,010 | -0,211 | 0,231 |
| | UKRCOAF-UKR | 0,392 | 0,009 | -0,374 | 0,391 |
| | UKR-INTER | 0,533 | 0,015 | -0,103 | 0,132 |
| | UKR-UKR | 0,253 | 0,010 | -0,030 | 0,051 |

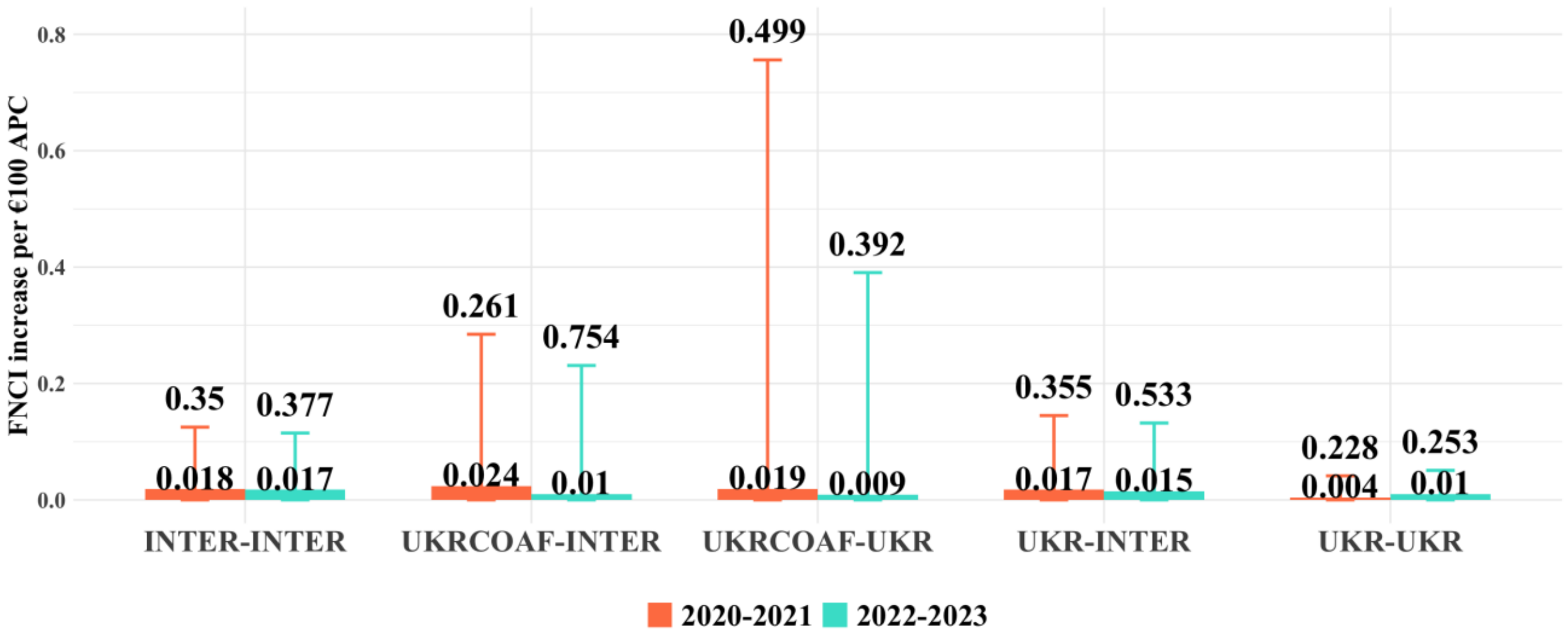

Figure 6. Baseline FNCI and effect of €100 APC by authorship type and

We examined the association between article processing charges (APC) and field-normalised citation impact (FNCI) using a linear regression model with log-transformed FNCI as the dependent variable (Model2). Predictors included APC (per €100), authorship type, period (2020–2021 vs 2022–2023), and all interactions. The model explained 18.6% of the variance in log(FNCI + 1) (adjusted $R^2 = 0.185$).

The top number above each bar (e.g., 0.554) represents the baseline FNCI from the regression model for each authorship type and period, corresponding to the predicted FNCI at €0 APC. Bars indicate the estimated increase in FNCI per additional €100 APC, while the numbers on the bars show the ΔFNCI for each group. Confidence intervals (error bars) represent 95% uncertainty around these estimates. Results suggest that higher APCs are associated with small but measurable increases in FNCI, with the magnitude of the effect varying by authorship pattern and period. For example, INTER-INTER papers had a baseline FNCI of 0.554 in 2020–2021, with an estimated increase of ΔFNCI = 0.028 per €100 APC, whereas UKRCOAF-INTER papers in 2022–2023 showed a larger increase of ΔFNCI = 0.045 per €100 APC.

Overall, APC effects are rather modest, and even large APC investments may yield only small citation boosts.

***Top Gold OA publishers***

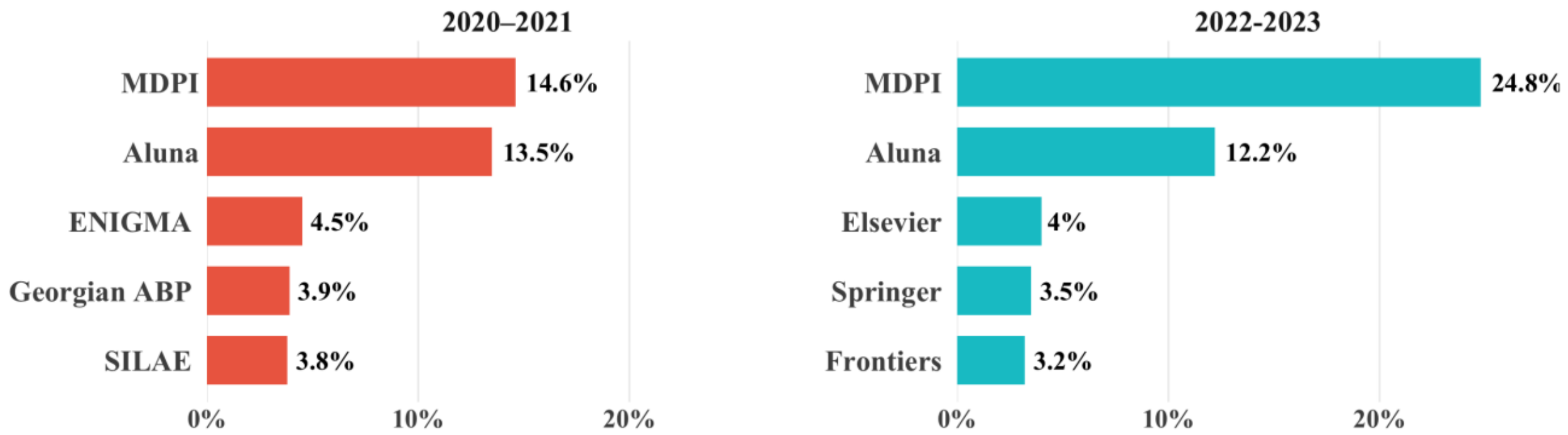

Figure 7. Top five publishers of Gold OA (percentage of total Gold OA in foreign journals).

Figure 7 lists the top five Gold OA publishers in 2020–2021 and 2022–2023. MDPI led in both periods, with its share increasing from 14.6% to 24.6%. It was followed by Aluna Publishing House, with most articles published in *Wiadomości Lekarskie* and some in *Polski Merkuriusz Lekarski*. In 2022–2023, Elsevier and Springer Nature entered the top five: Elsevier's output increased from 128 articles (1.8%) to 266 articles (4.0%), while Springer's rose from 179 articles (2.5%) to 236 articles (3.5%). Despite these changes, MDPI and Aluna remained the dominant Gold OA publishers for articles published by Ukrainian scholars.

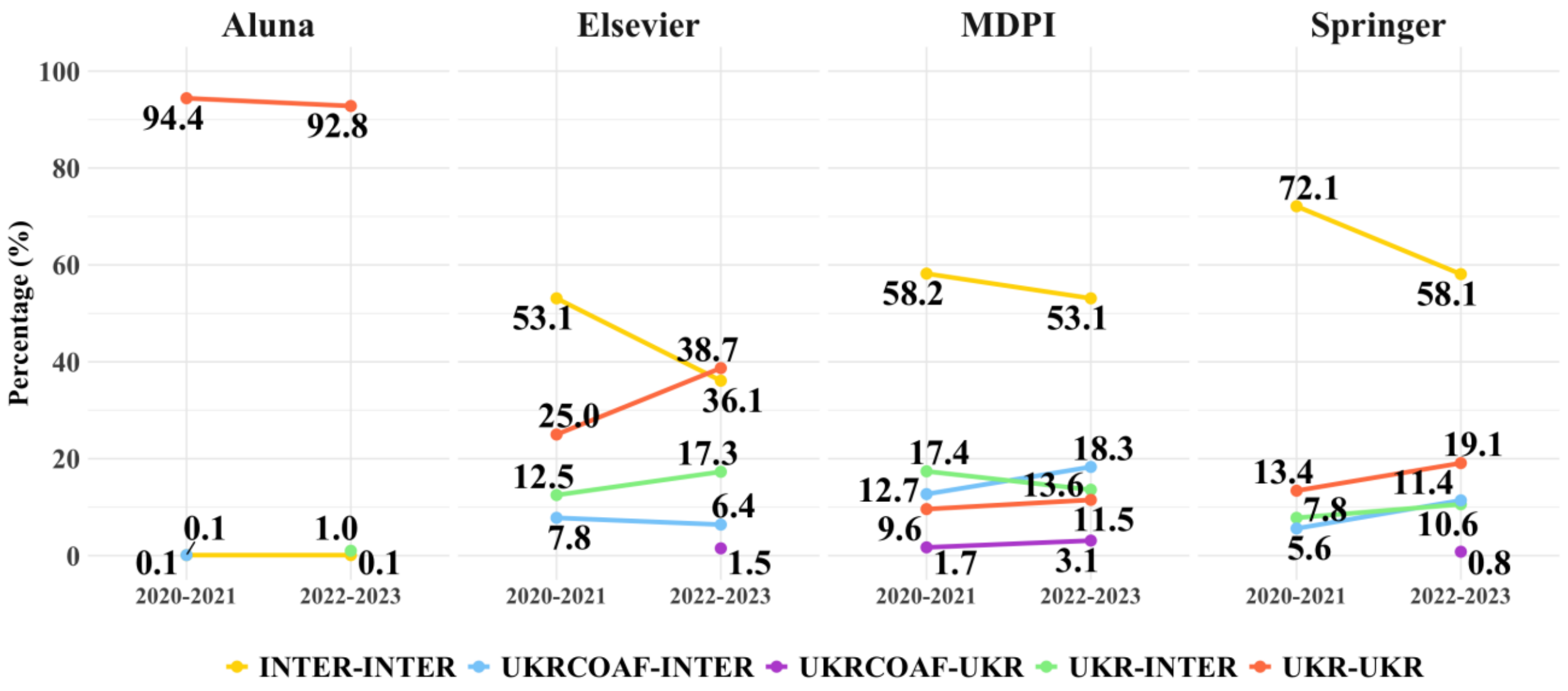

Figure 8. Distribution of articles across authorship types in journals published by Aluna, Elsevier, MDPI and Springer.

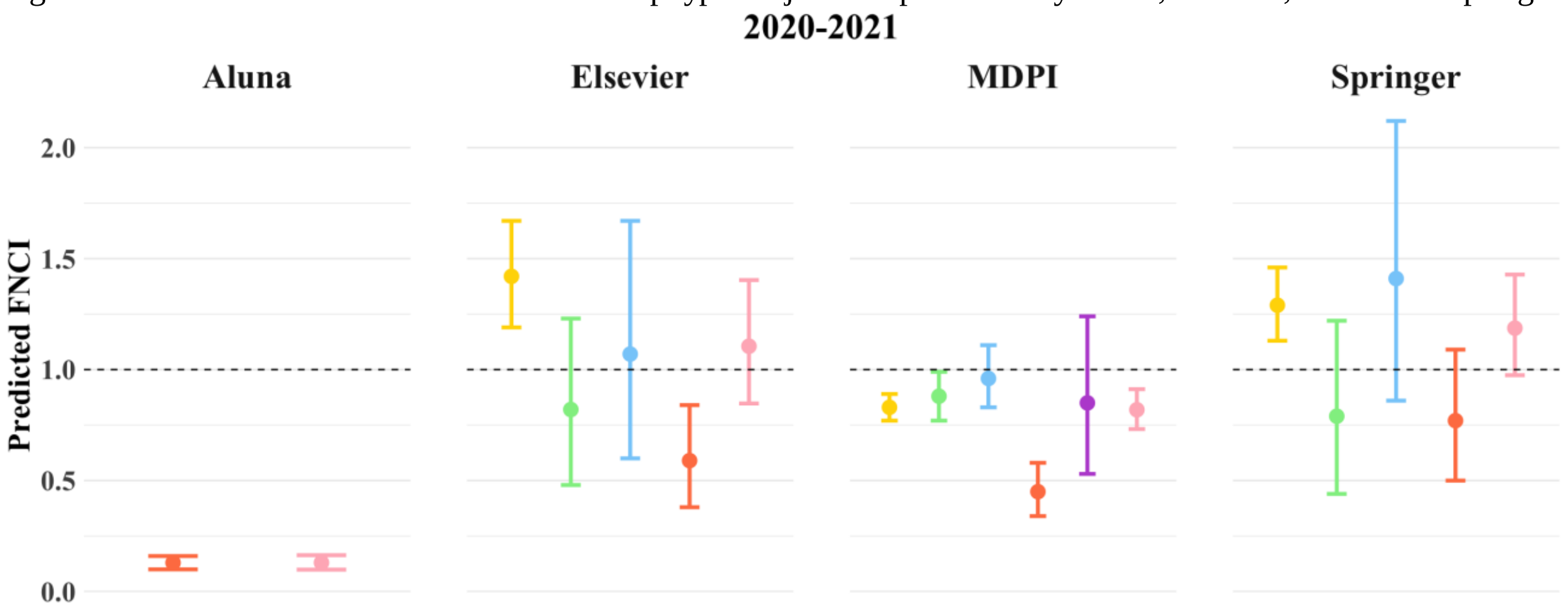

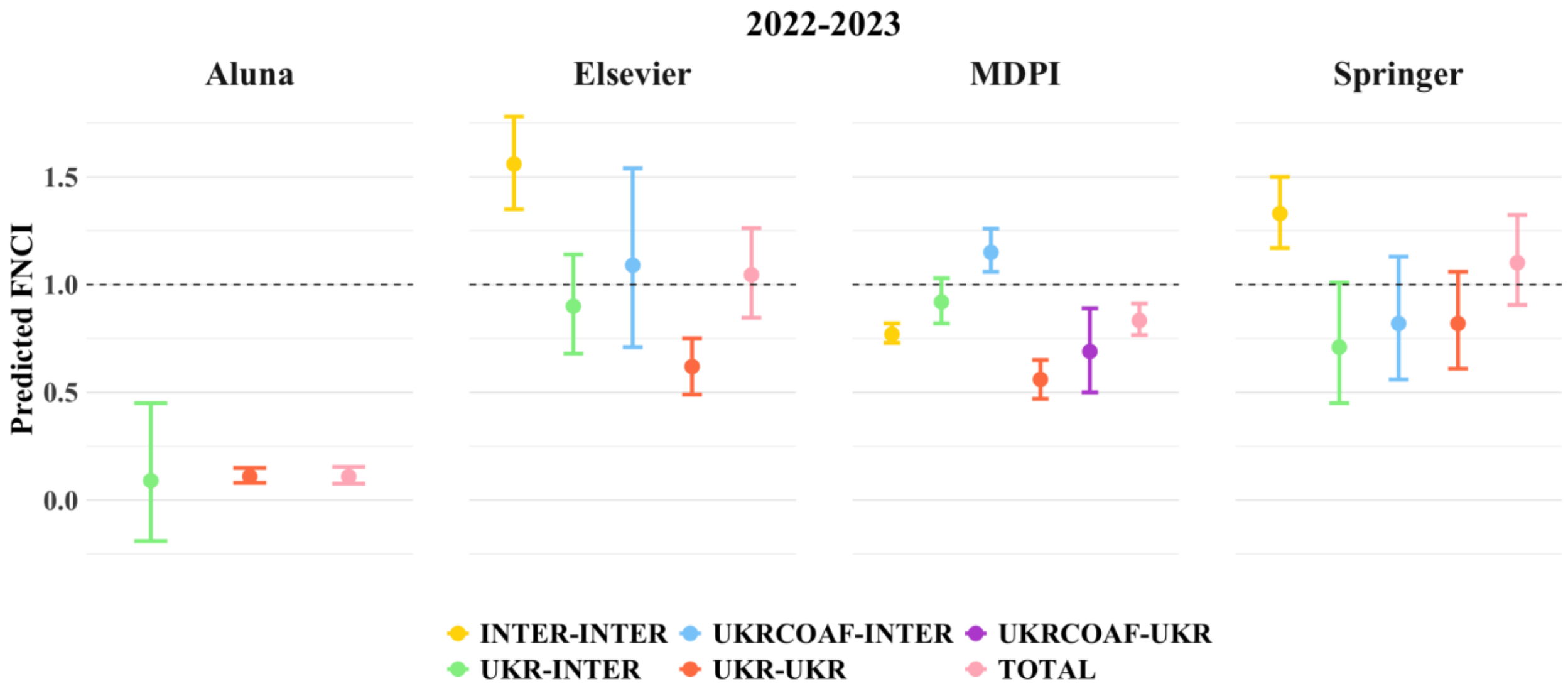

Fig 9 Predicted absolute FNCI values by publisher, authorship type and period

Figure 8 illustrates the distribution of authorship types across the publishers: Aluna Publishing House, Elsevier, MDPI, and Springer Nature. For Aluna Publishing House, more than 90% of articles in both periods were authored solely by Ukrainian scholars. This pattern may be attributed to the relatively lower APC of €600 compared with other publishers, as well as its reduction to €350 for Ukrainian scholars in 2022–2023. For the other publishers, INTER-INTER articles predominated over other authorship types. However, their share declined in 2022–2023, accompanied by an increase in other authorship categories. Specifically, for Elsevier the share of UKR-UKR articles for Elsevier rose sharply from 25.0% to 38.7%, exceeding the share of INTER-INTER articles. The increase in UKR-UKR articles was substantial for both Elsevier and Springer and more modest for MDPI. This pattern may reflect the impact of APC waivers provided by Elsevier and Springer to scholars affiliated with Ukrainian institutions. In contrast, for MDPI the largest increase was observed for UKRCOAF-INTER articles, whose share rose from 12.7% to 18.3%.

Model 3, a linear regression model predicting the logarithm of citation counts (log(FNCI + 1)) as a function of publication period, publisher, and authorship type, including interactions between period, publisher and authorship type, was estimated. The model explained 28.2% of the variance in log(FNCI + 1) (adjusted $R^2 = 0.282$).

Figure 9 presents the predicted absolute FNCI values derived from the model. Overall citation impact is lower for Aluna Publishing House (0.13 → 0.11) and MDPI (0.82 → 0.83) than for Elsevier (1.11 → 1.05) and Springer Nature (1.19 → 1.10), which exceed the world average (1.0). INTER-INTER articles, predominant in Elsevier, MDPI, and Springer, have the highest FNCI, surpassing 1.0 for Elsevier (1.42–1.56) and Springer (1.29–1.33), but remaining below 1.0 for MDPI (0.77–0.83). UKR-UKR articles show a similar trend, higher in Elsevier (0.59–0.62) and Springer (0.77–0.82) than MDPI (0.45–0.56). Overall citation impact of various authorship types differs across publishers.

*APC contribution vs. article share for Ukrainian-authored articles by publisher*

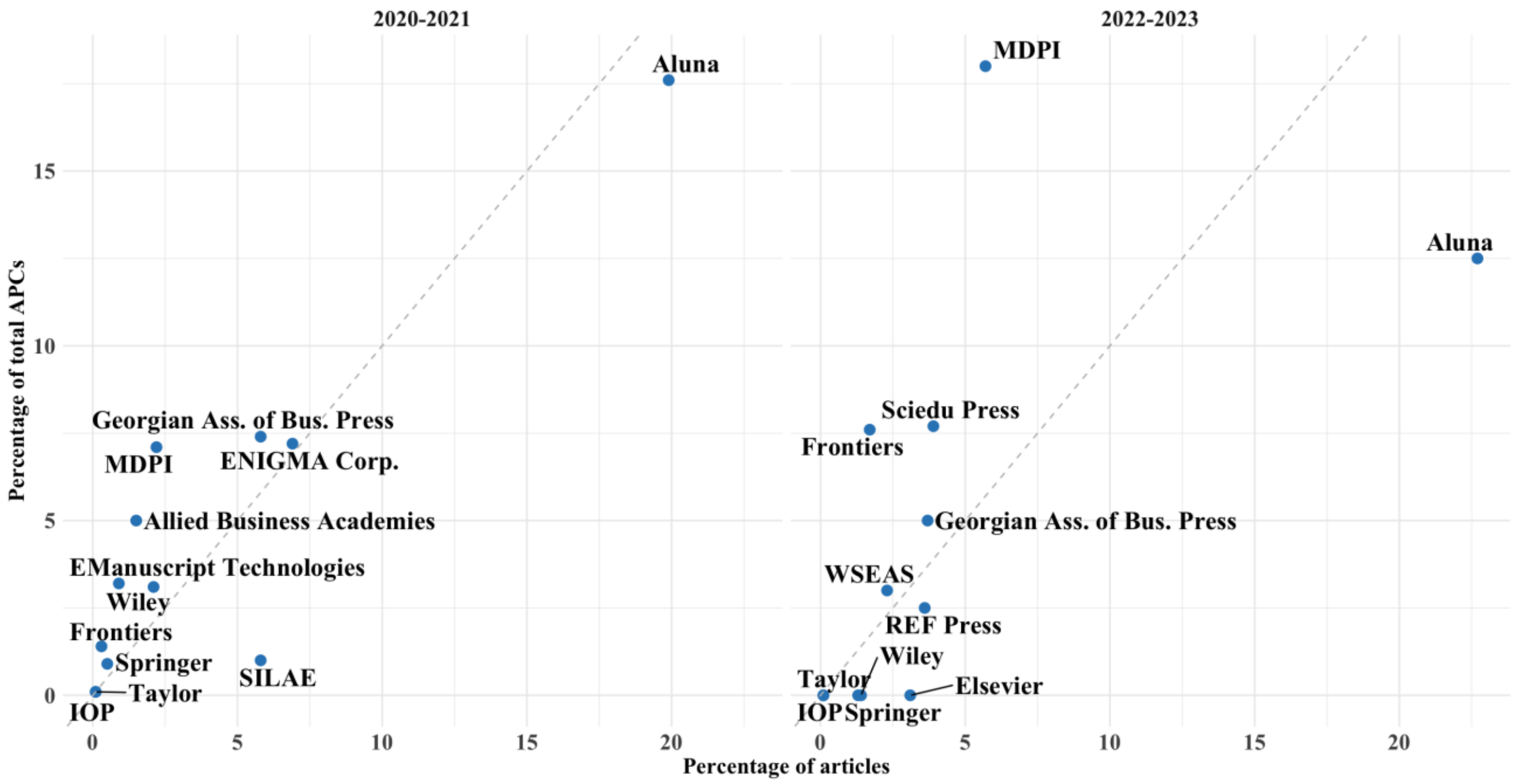

Figure 10. APC share versus article share by publisher for articles authored solely by Ukrainian authors (including publishers with ≥3% share and leading publishers offering waivers).

Figure 10 compares the percentage of APC share against the article share by publishers across two time periods. In both periods, Aluna Publishing House stands out with the highest percentages of both articles and APC. However, the percentage of articles exceeded the APC percentage—19.9% of articles versus 17.6% of APC in 2020–2021, and 22.7% of articles versus 12.5% of APC in 2022–2023. This widening gap likely reflects the discounts and APC reductions introduced by Aluna Publishing House to support Ukrainian scholars since the start of Russia's full-scale invasion of Ukraine. Meanwhile, MDPI saw the percentage of articles published rise from 2.2% to 5.7%, with the corresponding APC share increasing from 7.1% to 18.0%. The article share increased by approximately 2.6 times, whereas the APC share rose proportionally by about 2.5 times, indicating stable pricing relative to output. Publishers such as Frontiers, MDPI, Sciedu Press, WSEAS, and the Georgian Association of Business Press show disproportionately high APC shares relative to their article shares, suggesting higher costs for publishing in these venues. Despite waivers provided by major publishers like Springer and Elsevier, the percentage of articles published in their journals increased only marginally.

***APC costs for articles authored solely by Ukrainian scholars in foreign journals***

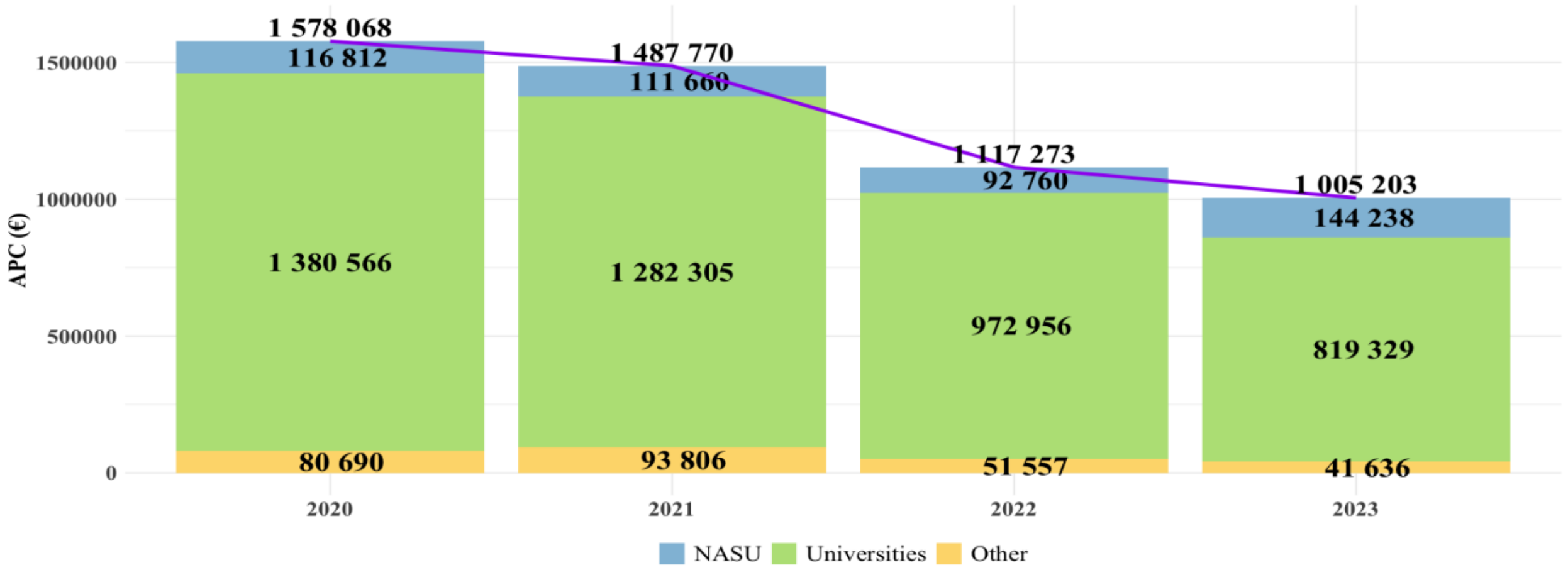

Figure 11. APC costs (euro) for articles authored solely by Ukrainian scholars in foreign journals.

Figure 11 shows that between 2020 and 2023, the total amount of APCs for articles authored solely by Ukrainian scholars in foreign journals decreased by 36%, from €1.6 million to €1.0 million. Over the same period, the number of articles authored solely by Ukrainian scholars declined by 30%. Notably, the total APCs paid by scholars affiliated with universities and other institutions fell, whereas APCs paid by scholars affiliated with the NASU increased in 2023. This divergence likely reflects the impact of waivers and discounts offered by publishers—factors not fully captured in the raw APC data.

*APC distribution relative to article output by discipline for articles authored solely by Ukrainian scholars*

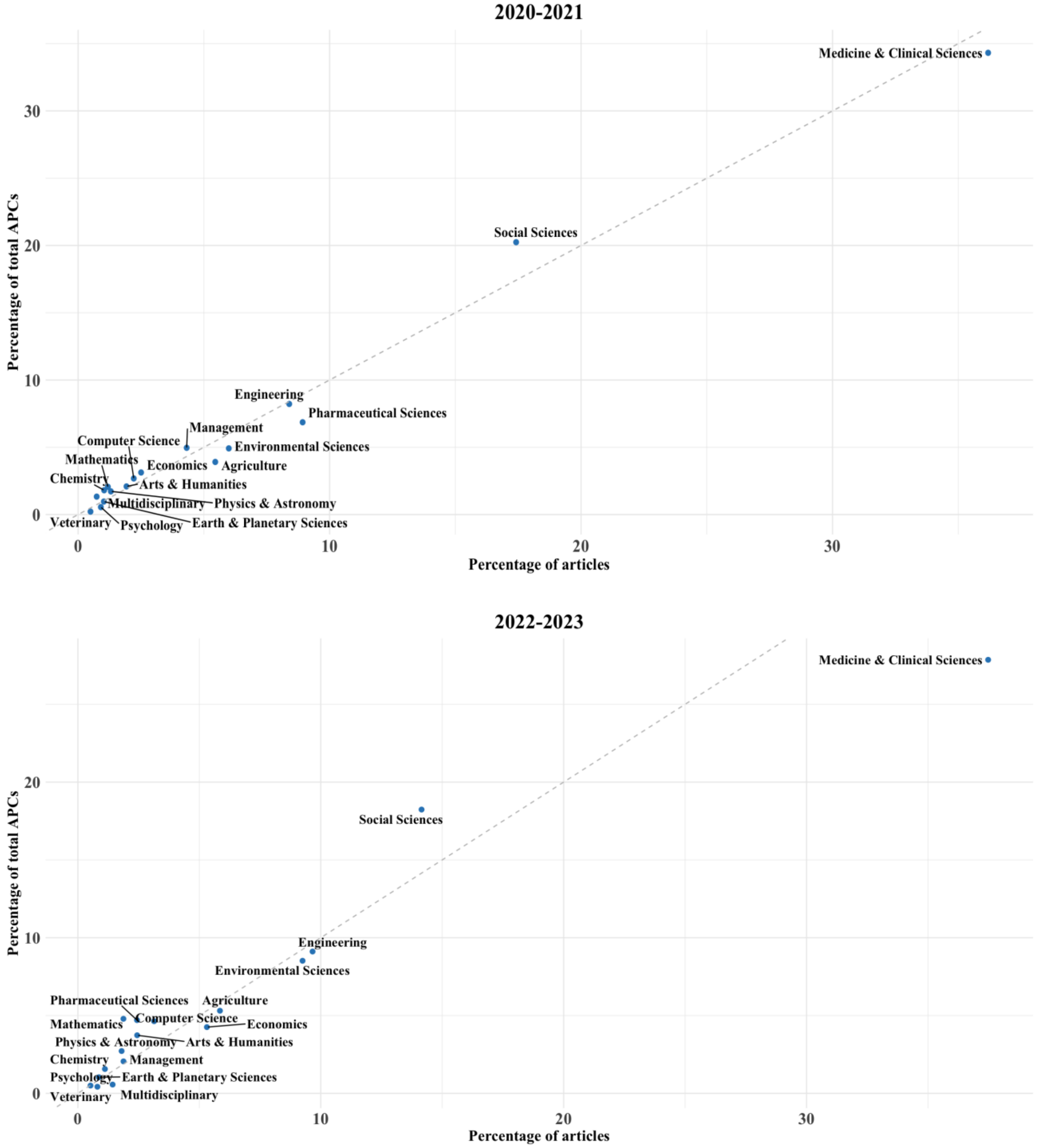

Figure 12. APC share against article share by discipline in articles authored solely by Ukrainian scholars.

Figure 12 compares the share of APCs with the share of articles across disciplines for the periods 2020–2021 and 2022–2023. In 2020-2021, the largest percentage of articles was in medicine & clinical sciences (36.2%), which consumed the largest share of APC costs (34.3%). In 2022-2023, medicine & clinical sciences preserved their dominance in the percentage of articles (37.5%) and the percentage of APC costs (27.9%). Considering that a large share of medical articles was published in Aluna journals (51.6% in 2020–2021 and 57.0% in 2022–2023), the reduction in APC costs can be partly attributed to Aluna lowering its APC fees from €600 to €350. The percentage of articles published in journals of publishers offering waivers to Ukrainian authors (Elsevier, Springer, SAGE, Taylor & Francis, Wiley) increased from 2.7% to 4.5%, with the number of such articles rising from 48 to 61. This could also contribute to a decline in the share of APC costs. In both periods, social sciences paid a higher APC share than their article share would predict: 17.4% against 20.2% in 2020-2021 and 14.1% against 18.2% in 2022-2023.

## Discussion and conclusions

### *Accuracy of Open Access metadata in Scopus*

Manual validation conducted in this study identified inconsistencies and occasional inaccuracies in the classification of open access publishing models, particularly Gold Open Access within Scopus metadata. Such misclassifications may compromise the reliability of bibliometric analyses and, consequently, the validity of research findings that depend on these data. These issues underscore the importance of rigorous metadata verification by indexing services. Given that researchers, institutions, and policymakers rely heavily on bibliographic databases for research evaluation, funding decisions, and strategic planning, improving metadata accuracy should be considered a priority.

### *Shifts in journal origin and authorship pattern*

Despite severe disruptions to research infrastructure and the economic downturn following Russia's full-scale invasion (OECD 2022; UNESCO 2024), APC-based Gold OA remained the dominant publishing model for Ukrainian research output in foreign journals between 2020 and 2023. Although major publishers such as Springer and Elsevier introduced APC waivers, the share of Ukrainian-led articles in these journals increased only marginally. Instead, publishers such as MDPI accounted for a larger proportion of Gold Open Access output. Consequently, publishing in Gold OA foreign journals under international leadership became more prevalent and was accompanied by a decline in articles authored solely by Ukrainian scholars.

### *Waivers for Gold OA and leading publishers*

As noted above, waivers and discounts offered by major publishers such as Springer and Elsevier had only a limited effect on publishing patterns, with MDPI and Aluna maintaining their dominance. While reduced article processing charges (APCs) may have supported Aluna's position, and the number of APCs waived by MDPI through reviewer vouchers remains unclear, both publishers appear to attract authors primarily due to their relatively low rejection rates and rapid publication timelines, despite the associated costs.

The prominence of the Polish publishing house Aluna aligns with prior studies showing that scholars from Central and Eastern Europe tend to publish extensively in journals based in neighboring countries (Grancay et al. 2017; Machacek and Srholec 2017; Pajic 2015; Pajic and Jevremov 2014), as well as in MDPI journals (Cernat 2024; Csomós and Farkas 2023; Sasvari and Urbanovics 2023; Kosmulski 2025). In the case of Aluna, comparatively low APCs likely further increase its attractiveness.

These patterns reflect a combination of structural constraints and strategic choices. On the one hand, Ukrainian scholars may have opted to pay APCs in journals with predictable acceptance outcomes rather than submit to Gold Open Access journals offering waivers (e.g., Elsevier and Springer), where acceptance is more competitive and the citation impact of articles is much higher. On the other hand, limited transparency and awareness of APC waiver policies, combined with the risk of policy changes during the review process, introduced additional uncertainty into an already fragile research environment.

In this context, stronger institutional support could have played a mitigating role. A more proactive role by the Ministry of Education and Science of Ukraine could have mitigated these challenges, for example by negotiating the terms and duration of APC waivers with major publishers. Notably, some publishers initially extended waivers to hybrid journals; however, in the absence of sustained national-level engagement, these provisions appear to have been discontinued. In 2026, Elsevier discontinued APC waivers for Ukrainian corresponding authors when co-authors are affiliated with institutions in high-income countries. Furthermore, in the context of research evaluation, the Ministry of Education and Science of Ukraine could have prioritized the quality of publications over quantity, discouraging scholars from spending state or personal funds on APCs during the ongoing conflict.

### *APC spending*

Prior work (Hladchenko, 2025) comparing access models found that articles in hybrid journals and preprints tend to have higher citation impact than both Gold OA and Diamond OA articles. This suggests that access model alone does not determine citation outcomes; factors such as article quality, authorship patterns and journal prestige likely play prominent roles (Glänzel and Schubert 2001; Glänzel 2001; Tahamtan et al. 2016; Abramo et al. 2019). In our APC effect analysis, this underscores why direct comparisons with Diamond OA—which typically has lower impact—are not appropriate without accounting for differences in journal indexing, subject coverage, and collaboration networks.

Our results indicate that overall Gold OA articles have citation impact below the world average and higher APCs are associated with modest increases in field-normalised citation impact (FNCI), with effects varying by authorship pattern and publication period. While statistically significant, these gains are relatively small and explain only a fraction of the variance in citation impact, suggesting that factors beyond APCs contribute to scholarly influence. This confirms that financial investment in publication does not guarantee higher impact (Maddi and Sapinho, 2022).

In Ukrainian scholarly publishing, medicine prevails as the discipline most represented by Gold OA publications in foreign journals (Hladchenko, 2025a). Medicine also accounts for the highest percentage of articles authored solely by Ukrainian scholars in foreign journals, as well as the largest share of APC expenditure. Most Gold OA articles were published in Aluna journals, which generally exhibited low citation impact. Notably, these articles often serve as prerequisites for awarding doctoral degrees and for promotion to associate professor and professor in Ukraine. A concerning pattern in the social sciences is that APC expenditures surpass the share of Ukrainian-authored articles, suggesting an inefficient allocation of publishing resources.

### *Concluding reflections and policy directions*

The findings reveal deep structural inequities in the APC-based Gold OA publishing model. Ukrainian scholars, like many in resource-limited settings, face systemic disadvantages as high publication fees divert scarce resources and limit publishing options. Waivers offer limited and inconsistent relief, while the shift toward fully Gold OA journals risks exacerbating these barriers. Without comprehensive reforms to reduce reliance on APCs and to balance incentives related to peer-review speed (Adam 2025), publishing costs, and disparities between scholars from various countries are likely to increase. Moreover, evaluation systems focused on output metrics have fostered a culture favouring rapid publication and APC payments over methodological rigour, even when high-quality, no-cost publishing exists. Shifting toward quality-focused assessments and investing in no-cost open-access publication venues can help reduce financial barriers and promote more inclusive knowledge production. Meaningful progress requires coordinated efforts from publishers, institutions, funders, and policymakers to create a more equitable and sustainable ecosystem. Only through systemic change can open access truly democratise scientific knowledge.

## Competing interests

The author declares no conflicts of interest relevant to the content of this article.

## Ethics approval

Not applicable.

## Consent for publication

Not applicable.

## Appendix

Table 1 Estimated coefficients from Model 1 examining the effects of publication period and authorship type on log-transformed FNCI (log(FNCI + 1))

| Term | Estimate | Lower 95% CI | Upper 95% CI | Standard Error | p.value |
|---|---|---|---|---|---|
| (Intercept) | 0.22 | 0.21 | 0.24 | 0.01 | <0.001 |
| 2022-2023 | 0.05 | 0.03 | 0.06 | 0.01 | <0.001 |
| INTER-INTER | 0.34 | 0.32 | 0.36 | 0.01 | <0.001 |
| INTER-UKR | 0.24 | 0.21 | 0.28 | 0.02 | <0.001 |
| INTER-UKRCOAF | 0.37 | 0.32 | 0.42 | 0.02 | <0.001 |
| UKR-UKRCOAF | 0.39 | 0.25 | 0.53 | 0.07 | <0.001 |
| 2022-2023:INTER-INTER | -0.04 | -0.07 | -0.01 | 0.02 | 0.012 |
| 2022-2023:INTER-UKR | 0.02 | -0.02 | 0.07 | 0.02 | 0.294 |
| 2022-2023:INTER-UKRCOAF | 0.03 | -0.03 | 0.09 | 0.03 | 0.311 |
| 2022-2023:UKR-UKRCOAF | -0.22 | -0.38 | -0.06 | 0.08 | 0.006 |

Table 2 Predicted FNCI by authorship type and period (with 95% confidence intervals)

| **Period** | Authorship type | Predicted log(FNCI + 1) | Predicted FNCI | Lower 95% CI | Upper 95% CI |
|---|---|---|---|---|---|
| 2020-2021 | UKR-UKR | 0.22 | 0.25 | 0.24 | 0.27 |
| 2020-2021 | INTER-INTER | 0.56 | 0.76 | 0.72 | 0.79 |
| 2020-2021 | INTER-UKR | 0.47 | 0.60 | 0.55 | 0.65 |
| 2020-2021 | INTER-UKRCOAF | 0.59 | 0.81 | 0.73 | 0.90 |
| 2020-2021 | UKR-UKRCOAF | 0.61 | 0.85 | 0.61 | 1.12 |
| 2022-2023 | UKR-UKR | 0.27 | 0.31 | 0.29 | 0.33 |
| 2022-2023 | INTER-INTER | 0.57 | 0.77 | 0.74 | 0.80 |
| 2022-2023 | INTER-UKR | 0.54 | 0.71 | 0.67 | 0.76 |
| 2022-2023 | INTER-UKRCOAF | 0.67 | 0.95 | 0.89 | 1.02 |
| 2022-2023 | UKR-UKRCOAF | 0.44 | 0.55 | 0.44 | 0.68 |
| 2020-2021 | TOTAL | 0.49 | 0.65 | 0.57 | 0.74 |
| 2022-2023 | TOTAL | 0.50 | 0.66 | 0.60 | 0.72 |

Table 3 Predicted FNCI by publisher, period, and collaboration type (with 95% confidence intervals and number of observations).

| **period** | **cor** | **Publisher** | **log_pred** | **FNCI_pred_abs** | **FNCI_lower** | **FNCI_upper** | **Number obs.** |
|---|---|---|---|---|---|---|---|
| | | Aluna | 0.08 | 0.08 | -0.52 | 1.44 | 1 |
| | | Elsevier | 0.88 | 1.42 | 1.19 | 1.67 | 68 |
| | | MDPI | 0.61 | 0.83 | 0.77 | 0.89 | 600 |
| | INTER-INTER | Springer | 0.83 | 1.29 | 1.13 | 1.46 | 129 |
| | | Aluna | 0.22 | 0.25 | -0.63 | 3.22 | |
| | | Elsevier | 0.6 | 0.82 | 0.48 | 1.23 | 16 |
| | | MDPI | 0.63 | 0.88 | 0.77 | 0.99 | 179 |
| | INTER-UKR | Springer | 0.58 | 0.79 | 0.44 | 1.22 | 14 |
| | | Aluna | 0.4 | 0.49 | -0.34 | 2.36 | 1 |
| | | Elsevier | 0.73 | 1.07 | 0.6 | 1.67 | 10 |
| | | MDPI | 0.67 | 0.96 | 0.83 | 1.11 | 131 |
| | INTER-UKRCOAF | Springer | 0.88 | 1.41 | 0.86 | 2.12 | 10 |
| | | Aluna | 0.12 | 0.13 | 0.1 | 0.16 | 899 |
| | | Elsevier | 0.47 | 0.59 | 0.38 | 0.84 | 32 |
| | | MDPI | 0.37 | 0.45 | 0.34 | 0.58 | 99 |
| | UKR-UKR | Springer | 0.57 | 0.77 | 0.5 | 1.09 | 24 |
| | | Aluna | -0.3 | -0.26 | -0.73 | 1.07 | |
| 2020-2021 | UKR-UKRCOAF | Elsevier | 0.78 | 1.19 | 0.35 | 2.54 | |
| | | MDPI | 0.62 | 0.85 | 0.53 | 1.24 | 18 |
| | | Springer | 0.45 | 0.58 | -0.16 | 1.95 | |
| | | Aluna | 0.1 | 0.14 | -0.43 | 1.65 | 901 |
| | | Elsevier | 0.69 | 1.02 | 0.6 | 1.59 | 126 |
| | | MDPI | 0.58 | 0.8 | 0.65 | 0.96 | 1027 |
| | TOTAL | Springer | 0.66 | 0.97 | 0.56 | 1.57 | 177 |
| | | Aluna | 0 | 0 | -0.56 | 1.25 | 1 |
| | | Elsevier | 0.94 | 1.56 | 1.35 | 1.78 | 96 |
| | | MDPI | 0.57 | 0.77 | 0.73 | 0.82 | 884 |
| | INTER-INTER | Springer | 0.85 | 1.33 | 1.17 | 1.5 | 137 |
| | | Aluna | 0.08 | 0.09 | -0.19 | 0.45 | 8 |
| | | Elsevier | 0.64 | 0.9 | 0.68 | 1.14 | 46 |
| | | MDPI | 0.65 | 0.92 | 0.82 | 1.03 | 227 |
| | INTER-UKR | Springer | 0.54 | 0.71 | 0.45 | 1.01 | 25 |
| | | Aluna | 0.02 | 0.02 | -0.76 | 3.34 | |
| | | Elsevier | 0.74 | 1.09 | 0.71 | 1.54 | 17 |
| | | MDPI | 0.77 | 1.15 | 1.06 | 1.26 | 305 |
| | INTER-UKRCOAF | Springer | 0.6 | 0.82 | 0.56 | 1.13 | 27 |
| | | Aluna | 0.11 | 0.11 | 0.08 | 0.15 | 1 |
| | | Elsevier | 0.48 | 0.62 | 0.49 | 0.75 | 96 |
| | | MDPI | 0.44 | 0.56 | 0.47 | 0.65 | 884 |
| | UKR-UKR | Springer | 0.6 | 0.82 | 0.61 | 1.06 | 137 |
| | | Aluna | -0.44 | -0.35 | -0.76 | 0.75 | 8 |
| | | Elsevier | 0.77 | 1.17 | 0.44 | 2.26 | 46 |
| | | MDPI | 0.52 | 0.69 | 0.5 | 0.89 | 227 |
| | UUKR-UKRCOAF | Springer | 0.41 | 0.5 | -0.15 | 1.67 | 25 |
| | | Aluna | -0.04 | -0.03 | -0.44 | 1.19 | 768 |
| | | Elsevier | 0.71 | 1.07 | 0.74 | 1.49 | 266 |
| | | MDPI | 0.59 | 0.82 | 0.72 | 0.93 | 1658 |
| 2022-2023 | TOTAL | Springer | 0.6 | 0.84 | 0.53 | 1.27 | 236 |